\documentclass[aps,prd,superscriptaddress,showpacs,amsmath,amssymb]{revtex4}
\usepackage{graphicx}
\usepackage{epsf}

\usepackage[]{latexsym}
\usepackage{bm}

\newcommand{\be}{\begin{equation}}\newcommand{\ee}{\end{equation}}
\newcommand{\bea}{\begin{eqnarray}}\newcommand{\eea}{\end{eqnarray}}
\newcommand{\brr}{\begin{array}}\newcommand{\err}{\end{array}}
\newcommand{\bit}{\begin{itemize}}\newcommand{\eit}{\end{itemize}}
\newcommand{\ben}{\begin{enumerate}}\newcommand{\een}{\end{enumerate}}

\newcommand{\ba}{\begin{array}}
\newcommand{\ea}{\end{array}}

\def\lan{\langle}
\def\lf{\left}

\def\non{\nonumber}\def\ran{\rangle}

\def\ri{\right}
\def\al{\alpha}

\def\te{\theta}
\def\si{\sigma}
\def\om{\omega}

\def\1{{_{1}}}\def\2{{_{2}}}

\def\noHe0{:\;\!\!\;\!\!:H_e(0):\;\!\!\;\!\!:}
\def\noHm0{:\;\!\!\;\!\!:H_\mu(0):\;\!\!\;\!\!:}

\def\lan{\langle}
\def\lf{\left}

\def\non{\nonumber}
\def\ran{\rangle}

\def\ri{\right}

\def\al{\alpha}

\def\te{\theta}

\def\si{\sigma}
\def\om{\omega}

\def\1{{_{1}}}\def\2{{_{2}}}

\begin{document}

\title{Vacuum condensates, flavor mixing and spontaneous supersymmetry breaking}

\author{ Antonio Capolupo${}^{\natural}$\footnote{Corresponding author, e-mail address: capolupo@sa.infn.it}}

 \affiliation{${}^{\natural}$ Dipartimento di Ingegneria Industriale,
  Universit\'a di Salerno, Fisciano (SA) - 84084, Italy}
\author{Marco Di Mauro}


\begin{abstract}

Spontaneous supersymmetry (SUSY) breaking is revealed in all phenomena  in which vacuum condensates are physically relevant.
The dynamical breakdown of SUSY is generated by
  the   condensates themselves, which lift the zero point energy. Evidence is presented in the case of the Wess-Zuimino model, and the flavor mixing case is treated in detail.

\end{abstract}

\maketitle

\section{Introduction}

Supersymmetry (SUSY) has had an enormous impact on physics in the last four decades, since it was first proposed in 1971 \cite{Golfand:1971iw}. From the purely theoretical side, we mention  the many instances of supersymmetric quantum field theories which are highly controllable. These taught us a great deal about previously poorly understood subjects, such as strongly coupled dynamics of four dimensional quantum field theories and duality, and also had a huge influence on mathematics.
From the phenomenological point of view,  SUSY has had even bigger effects. In fact, the assumption that SUSY is a fundamental symmetry of nature gives a very natural solution to the hierarchy problem, that is of the naturalness of a fundamental scalar Higgs field, which is impossible to give in the framework of the Standard Model. This spurred an enormous activity on the construction of phenomenological models which display SUSY.

The influence of SUSY has been important also on experimental physics, since much of the research has been geared at the detection of the superpartners which according to SUSY are associated with the ordinary particles we know. Despite the enormous efforts devoted to this task, no evidence of the existence of superpartners has ever been given.
 The fact that the superpartners are not degenerate with ordinary particles (otherwise they would have been detected) means that, if SUSY exists as a fundamental symmetry, it must be spontaneously broken at the scales we can currently probe. As a consequence a large part of the above mentioned research activity has been devoted to the study of SUSY breaking. In this paper we concentrate on the spontaneous breaking of SUSY, leaving aside the possibility of explicit breaking which has also been much considered in the literature.

The first models exhibiting the spontaneous breaking of SUSY were proposed in \cite{Fayet:1974jb,O'Raifeartaigh:1975pr}. In these models the breaking is classical, i.e. it occurs at tree level. Another possibility is that the breaking is dynamical, i.e. it is triggered by nonperturbative quantum effects. This most interesting possibility has been first discussed in \cite{Witten:1981nf} (see \cite{Shadmi:1999jy} for a review).

It is very important to stress that, even if SUSY will not be recognized to be a fundamental symmetry of nature, it can be realized as an emergent symmetry, e.g. in condensed matter systems \cite{Yue Yang}. The following discussion is very general and applies to both fundamental and emergent SUSY.

In an apparently detached line of study, the r\^{o}le played by vacuum condensates in many Quantum Field Theoretical phenomena has been analyzed. Examples of systems characterized by the presence of condensates include QFT in  external fields like Unruh \cite{Unruh:1976db} and  Schwinger effects \cite{Schwinger:1951nm}, condensed matter physics (BCS theory of superconductivity \cite{Bardeen:1957mv}, graphene physics \cite{Iorio:2010pv}, Thermo Field Dynamics \cite{Takahasi:1974zn}), particle physics and cosmology (flavor mixing \cite{Blasone:1995zc,Blasone:2002jv,Blasone:2001du,Blasone:2005ae,Mishchenko1,JI1,JI2,Capolupo:2004pt}, dark energy \cite{Capolupo:2006et,Capolupo:2008rz,Capolupo:2007hy}),  and quantization of dissipative systems \cite{Celeghini:1991yv}.  In all cases, vacuum condensates can be effectively described by using Bogoliubov transformations. The specific details of the mechanism or of the field that induces the condensate are contained in the coefficients of such transformations.

The purpose of this paper is to stress the connection between the two issues outlined above. In fact, it has been recently proposed \cite{Capolupo:2012vf} that, in a supersymmetric context, vacuum condensates, generated in phenomena like the above listed ones, provide a new mechanism of spontaneous SUSY breaking. A particulary interesting system from the phenomenological point of view is represented by flavor mixing \cite{Capolupo:2010ek} (see also \cite{Mavromatos:2010ni}),  which appears in both the hadronic and leptonic sectors of the Standard Model.

In the following we study a system whose  Lagrangian is invariant under SUSY, namely the free Wess--Zumino model. We then implement the vacuum condensation effects by means of a Bogoliubov transformation, which acts simultaneously and with the same parameters on the bosonic and on the fermionic degrees of freedom, in order not to break SUSY explicitly. We conjecture that in this situation SUSY will be spontaneously broken, since the presence of the condensates shifts the vacuum energy density to a nonvanishing value. As well known, indeed, this is a sufficient condition for the spontaneous breaking of SUSY \cite{Witten:1981nf}.
The SUSY breaking revealed considering the free Wess--Zumino (WZ) model should give a qualitative understanding of the behavior of more complicate systems.

The paper is organized as follows. In Sec.2, we discuss the effects of Bogoliubov transformations in QFT on vacuum energy and we show the SUSY breaking for free WZ model induced by condensates. In Sec.3, we treat the specific case of particle mixing, and Sec.4 is devoted to the discussion of our results and to conclusions.

\section{Vacuum condensate, WZ model and SUSY breaking}

In the following we summarize different aspects of Bogoliubov transformations in the context of quantum field theory \cite{Umezawa:1993yq} and we study the effects of vacuum condensates in the free Wess-Zumino model \cite{Wess:1973kz}. We start by treating the bosonic case. The fermion case is analogous and will be considered when the Wess--Zumino Lagrangian will be analyzed explicitly.

Any physical degree of freedom of a bosonic field is described by a set of ladder operators $a_{\mathbf{k}}$ with canonical commutation relations (CCRs), $ [a_{\mathbf{k}}, a^{\dagger}_{\mathbf{p}}]=\delta^{3}(\mathbf{k}-\mathbf{p})\,,$
with all other commutators vanishing. To such a set a vacuum $|0\rangle$ is associated, defined by $a_{\mathbf{k}}|0\rangle=0$, and a Fock space is given as an irreducible representation of the CCRs algebra.

A general Bogoliubov transformation  acts on these bosonic modes as
$\tilde{a}_{\mathbf{k}}(\xi) = U_{\mathbf{k}}(\xi) \, a_{\mathbf{k}} - V_{\mathbf{k}}(\xi) \,  a^{\dagger}_{\mathbf{k}}\,, $
with the condition, $|U_{\mathbf{k}} |^2 - |V_{\mathbf{k}} |^2 = 1\,$, which ensures the invariance of the CCRs.
The parameter $\xi$, which in general depends on the mode, is associated to the physics which is implemented by the transformation, e.g. the temperature in the TFD case, or the acceleration in the Unruh case.
The Bogoliubov transformation can be expressed by means of a generator $J(\xi)$ in the form
$ \tilde{a}_{\mathbf{k}}(\xi) = J^{-1}(\xi)\,  a_{\mathbf{k}}\,  J(\xi)\, $, with
$ J(\xi) = \exp \lf[\frac{i}{2}\sum_{\mathbf{k}} \xi_{\mathbf{k}}\lf(a_{\mathbf{k}}^2 - (a_{\mathbf{k}}^{\dagger})^2\ri)\ri]\,$.
The transformed modes $\tilde{a}_{\mathbf{k}}(\xi)$ annihilate a state $|\tilde{0}(\xi)\rangle$, i.e. $\tilde{a}_{\mathbf{k}}(\xi)|\tilde{0}(\xi)\rangle=0$, which is related to the vacuum $|0\rangle$ by
$ |\tilde{0}(\xi)\rangle = J^{-1} (\xi)|0\rangle\,.$
Such a state is called itself a vacuum, for the following reason. The generator $J$ is a unitary operator if $\mathbf{k}$ assumes a discrete  range of values; in this case, the Fock spaces built on the  states $|0\rangle $ and $|\tilde{0}(\xi)\rangle$ are equivalent. On the other hand if $\mathbf{k}$ assumes a continuous infinity of values, which is what happens in the case of QFT, $J$ is not a unitary operator any more. This implies that the state $|\tilde{0}(\xi) \rangle$ cannot be expressed as a superposition of vectors belonging to the Fock space built over $|0\rangle$. Rather, it is a new vacuum, over which a whole new Fock space can be built by using the tilded creation operators. The two Fock spaces are unitarily inequivalent and describe different physical situations.

The above described Bogoliubov transformation acts on the modes of a single field. A second type of Bogoliubov transformation mixes the modes of two fields, and its effects are essentially the same as in the single field case. Such is for example the one which is involved in the case of particle mixing, as we shall see later.

Let us now consider the free Wess--Zumino model.
The Lagrangian is given by
\bea \label{WS}
\mathcal{L} = \frac{i}{2} \,\bar{\psi}\gamma_{\mu}\partial^{\mu}\psi + \frac{1}{2}\,\partial_{\mu}S\partial^{\mu}S + \frac{1}{2}\,\partial_{\mu}P\partial^{\mu}P - \frac{m}{2} \, \bar{\psi}\psi - \frac{m^2}{2}\, (S^2 + P^2),
\eea
where $\psi$ is a Majorana spinor field, $S$ is a scalar field and $P$ is a pseudoscalar field. This Lagrangian is invariant under supersymmetry transformations \cite{Wess:1973kz}.
By quantizing the fields in the usual way and denoting with $\alpha^r_{\mathbf{k}}$, $b_{\mathbf{k}}$ and $c_{\mathbf{k}}$ the annihilator of
$\psi$, $S$ and $P$ respectively, we define the vacuum $|0\rangle=|0\rangle_{\psi}\otimes|0\rangle_S \otimes |0\rangle_P$  as the state annihilated by such annihilators.

Now we perform a Bogoliubov transformation on the ladder operators corresponding to all three fields,
\bea\label{Bog1}
\tilde{\alpha}^r_{\mathbf{k}}(\xi, t) &=& U^{\psi}_{\mathbf{k}} \, \alpha^r_{\mathbf{k}}(t) + V^{\psi}_{-\mathbf{k}} \, \alpha^{r\dagger}_{-\mathbf{k}}( t)\,, \non
\\
\tilde{\alpha}^{r\dagger}_{-\mathbf{k}}(\xi,  t) &=& U^{\psi *}_{-\mathbf{k}} \, \alpha^{r\dagger}_{-\mathbf{k}}(t) + V^{\psi *}_{ \mathbf{k}} \, \alpha^{r}_{\mathbf{k}}(t)\,,
\\[2mm]\non
\tilde{b}_{\mathbf{k}}(\eta,  t) &=& U^{S}_{\mathbf{k}} \, b_{\mathbf{k}}(t) - V^{S}_{-\mathbf{k}}  \,b^{\dagger}_{-\mathbf{k}}(t)\,,
\\\label{Bog2}
\tilde{b^{\dag}}_{-\mathbf{k}}(\eta,  t) &=& U^{S *}_{-\mathbf{k}} \, b^{\dag}_{-\mathbf{k}}(t) - V^{S *}_{\mathbf{k}}  \,b_{\mathbf{k}}(t)\,,
\\[2mm]\non
\tilde{c}_{\mathbf{k}}(\eta,  t) &=& U^{P}_{\mathbf{k}} \, c_{\mathbf{k}}(t) - V^{P}_{-\mathbf{k}}  \,c^{\dagger}_{-\mathbf{k}}(t)\,,
\\\label{Bog3}
\tilde{c^{\dag}}_{-\mathbf{k}}(\eta,  t) &=& U^{P *}_{-\mathbf{k}} \, c^{\dag}_{-\mathbf{k}}(t) - V^{P *}_{\mathbf{k}}  \,c_{\mathbf{k}}(t)\,,
\eea
Since the Bogoliubov coefficients of scalar and pseudoscalar bosons are the same, $U^{S}_{\mathbf{k}} =U^{P}_{\mathbf{k}} $ and $V^{S}_{\mathbf{k}} =V^{P}_{\mathbf{k}} $, we denote them as $U^{B}_{\mathbf{k}} $ and $V^{B}_{\mathbf{k}} $, respectively. The constraints satisfied are $ U^{\psi}_{\mathbf{k}} = U^{\psi}_{-\mathbf{k}}$,  $ V^{\psi}_{\mathbf{k}} = -V^{\psi}_{-\mathbf{k}},$ and $ |U^{\psi}_{\mathbf{k}}|^2 + |V^{\psi}_{\mathbf{k}}|^2 = 1\,$  for fermions and
$ U^{B}_{\mathbf{k}} = U^{B}_{-\mathbf{k}}$, $ V^{B}_{\mathbf{k}} = V^{B}_{-\mathbf{k}}$, and $ |U^{B}_{\mathbf{k}}|^2 - |V^{B}_{\mathbf{k}}|^2 = 1$ for bosons. Thus, the forms of the coefficients are
$U^{\psi}_{\mathbf{k}} =e^{i\phi_{1\mathbf{k}}}\cos\xi_{\mathbf{k}}(\zeta)$, $ V^{\psi}_{\mathbf{k}} =e^{i\phi_{2\mathbf{k}}}\sin\xi_{\mathbf{k}}(\zeta)$, $U^{B}_{\mathbf{k}} = e^{i\gamma_{1\mathbf{k}}}\cosh\eta_{\mathbf{k}}(\zeta)$, $V^{B}_{\mathbf{k}} =e^{i\gamma_{2\mathbf{k}}}\sinh\eta_{\mathbf{k}}(\zeta)$,
respectively, where $\zeta$ is the  parameter which controls the physics underlying the  transformation. We neglect the phases $\phi_{i\mathbf{k}}$, $\gamma_{i\mathbf{k}}$, $i=1,2$, since they are irrelevant in the following.

At any time $t$ the transformations (\ref{Bog1})--(\ref{Bog3}) can be written as:
$
\tilde{\alpha}^r_{\mathbf{k}}( \xi,  t) = J^{-1} (\xi,\eta,  t)\,\alpha^r_{\mathbf{k}}(t) J(\xi,\eta,  t)\,,
$
and similar relations for the other  operators, where the generator is
$
J(\xi,\eta,   t) = J_{\psi}(\xi,  t) J_{S}(\eta,  t) J_{P}(\eta,  t)\,,
$
with
$
J_{\psi} = \exp \lf[\frac{1}{2}\int d^{3} \mathbf{k} \, \xi_{\mathbf{k}}(\zeta)\lf(\alpha^r_{\mathbf{k}}(t) \alpha^r_{-\mathbf{k}}(t)  - \alpha^{r \dagger}_{-\mathbf{k}}(t) \alpha_{ \mathbf{k}}^{r \dagger}(t)  \ri)\ri]
$,
$
J_{S} = \exp \lf[-i\int d^{3} \mathbf{k}\, \eta_{\mathbf{k}}(\zeta)\lf(b_{\mathbf{k}}(t) b_{-\mathbf{k}}(t)  - b_{-\mathbf{k}}^{\dagger}(t) b_{ \mathbf{k}}^{\dagger}(t)  \ri)\ri]
$ and
$
J_{P} = \exp \lf[-i\int d^{3} \mathbf{k}\, \eta_{\mathbf{k}}(\zeta)\lf(c_{\mathbf{k}}(t) c_{-\mathbf{k}}(t)  - c_{-\mathbf{k}}^{\dagger}(t) c_{ \mathbf{k}}^{\dagger}(t)  \ri)\ri].
$
The new annihilators define the tensor product vacuum
$|\tilde{0}(   t)\rangle=|\tilde{0}(   t)\rangle_{\psi}\otimes |\tilde{0}(  t)\rangle_{S}\otimes |\tilde{0}(   t)\rangle_{P}$, where
$
|\tilde{0}(  t)\rangle_{\psi} = J^{-1}_{\psi}(\xi,  t)|0\rangle_{\psi} $,
$ |\tilde{0}(  t)\rangle_{S} = J^{-1}_{S}(\eta,  t)|0\rangle_{S}$ and
$  |\tilde{0}(   t)\rangle_{P} = J^{-1}_{P}(\eta,  t)|0\rangle_{P}
$.
Then
$
|\tilde{0}(   t)\rangle = J^{-1}(\xi,\eta,  t)|0\rangle\,.
$
The vacuum
$|\tilde{0}( t)\rangle$  is the relevant, physical vacuum for the systems listed above when they are studied in a supersymmetric context. It has the nontrivial structure of a condensate of couples of particles and antiparticles as can be seen explicitly by looking at the condensation densities of fermions and bosons,
$\langle\tilde{0}( t)| \alpha^{r\dagger}_{\mathbf{k}} \alpha^r_{\mathbf{k}} |\tilde{0}(  t)\rangle  = |V_{\mathbf{k}}^{\psi}|^2\,$, and
$\langle\tilde{0}(   t)| b^{\dagger}_{\mathbf{k}} b_{\mathbf{k}} |\tilde{0}(   t)\rangle  =  \langle\tilde{0}( t)| c^{\dagger}_{\mathbf{k}} c_{\mathbf{k}} |\tilde{0}( t)\rangle = |V_{\mathbf{k}}^{B}|^2$. Such vacuum condensates are responsible for the energy density of $|\tilde{0}( t)\rangle$ being different from zero.

In fact, denoting with $H= H_{\psi} +  H_B $  the free Hamiltonian  corresponding to the Lagrangian (\ref{WS}),  where $H_B = H_S + H_P$, one has $
\langle\tilde{0}(  t)| H_{\psi} |\tilde{0}(  t)\rangle = - \int \; d^3\mathbf{k}\; \omega_{\mathbf{k}} \,(1- 2 |V^{\psi}_{\mathbf{k}}|^2)\,,$ for fermions and
$ \langle\tilde{0}(   t)| H_B |\tilde{0}(  t)\rangle = \int\; d^3\mathbf{k}\; \omega_{\mathbf{k}} (1 + 2 |V^{B}_{\textbf{k}}|^2)\,,$
for bosons. Then the expectation value of the Hamiltonian $H$ on  $|\tilde{0}( t)\rangle$ is \cite{Capolupo:2012vf}
\bea\label{Ht}
\langle\tilde{0}(  t)| H |\tilde{0}(  t)\rangle = 2 \int\; d^3\mathbf{k}\; \omega_{\mathbf{k}} (|V^{\psi}_{\textbf{k}}|^2 + |V^{B}_{\textbf{k}}|^2)\,,
\eea
which is different from zero and positive because of the presence of the fermion  and boson
condensates, both of which lift  the vacuum energy by a positive amount. This result,  together with the fact that the true vacuum is the transformed one $|\tilde{0}(  t)\rangle$, implies, in  supersymmetric context, the spontaneous breaking of SUSY induced by phenomena in which condensation takes place.

The above result, Eq.(\ref{Ht}), holds for disparate physical phenomena; the explicit form of the Bogoliubov coefficients
$V^{\psi}_{\textbf{k}}$ and $V^{B}_{\textbf{k}}$ specifies the particular system.
For example, in the case of Thermo Field Dynamics, the  parameter  $\zeta$ is the temperature, the physical vacuum is the thermal one, and the result is that SUSY is spontaneously broken at any nonzero temperature, as is well known \cite{Das:1997gg,Buchholz:1997mf}.

\section{Flavor mixing and SUSY breaking}

Now we consider explicitly the specific case of flavor mixing  because of the phenomenological interest of this system. Our starting point is the Wess--Zumino lagrangian for two massive free chiral supermultiplets
\bea \label{WS2}
\mathcal{L} = \sum_{i=1}^2\lf\{\frac{i}{2} \bar{\psi}_i\gamma_{\mu}\partial^{\mu}\psi_i + \frac{1}{2}\partial_{\mu}S_i\partial^{\mu}S_i + \frac{1}{2}\partial_{\mu}P_i\partial^{\mu}P_i - \frac{m}{2}  \bar{\psi}_i\psi_i - \frac{m^2}{2} (S_i^2 + P_i^2)\ri\},
\eea
with $\psi_i$ denoting two free Majorana fermions, $S_i$ two free real scalars and  $P_i$ two free real pseudoscalars ($i=1,2$).
We assume that $m_1\neq m_2$, since this is a necessary condition to have nontrivial mixing. It is clear that both the $1$ and $2$ sectors of the above Lagrangian are separately invariant under SUSY transformations.

Consider now the mixing transformations for the Lagrangian (\ref{WS2}):
\bea\label{mixing-transf}
\psi_f= U \psi, \qquad S_f = U S,\qquad P_f= U P
\eea
where $U = \lf(\begin{array}{cc} \cos\theta&\sin\theta\\ -\sin\theta &\cos \theta\end{array}\ri)$, $\psi=(\psi_1, \psi_2)^T$, $\psi_f=(\psi_a, \psi_b)^T = (\cos\theta \,\psi_1 + \sin\theta\,\psi_2, -\sin\theta\,\psi_1 + \cos\theta\,\psi_2)^T $ , etc.
The Lagrangian (\ref{WS2}) takes then the form
\bea\label{MixLagrangian}
\mathcal{L}=\frac{i}{2} \bar{\psi}_f( \not\!\partial + M)\psi_f + \frac{1}{2}\partial_{\mu}S_f \partial^{\mu}S_f
-\frac{1}{2} S_f^{T}\, M^2 \,S_f +\frac{1}{2} \partial_{\mu}P_f \,\partial^{\mu}P_f -\frac{1}{2} P_f^{T} \,M^2 \,P_f \,,
\eea
with $M=\lf(\begin{array}{cc} m_a&m_{ab}\\ m_{ab} &m_b\end{array}\ri)$, where $m_a = m_1\cos^2\theta + m_2\sin^2\theta$, $m_b = m_1\sin^2\theta + m_2\cos^2\theta$, and $m_{ab}=(m_2-m_1)\sin\theta\cos\theta$. This Lagrangian is invariant under the same SUSY transformations as the unmixed one, but this time referring to the mixed field. A careful discussion of this fact can be found in \cite{Iorio:2010pv}.

Let us now proceed as in the previous section.
We quantize the fields  $\psi_{i}$, $S_{i}$ and $P_{i}$  and we express the mixing transformations
(\ref{mixing-transf})  in terms of generators as
$
\psi_{\si}(x)  \equiv  G^{-1}_{\psi }(\te) \; \psi_{i}(x)\; G_{\psi }(\te)
$,
$
S_{\si}(x)  \equiv  G^{-1}_S(\te) \; S_{i}(x)\; G_S(\te)
$ and
$
P_{\si}(x) \equiv  G^{-1}_{P}(\te) \; P_{i}(x)\; G_{P}(\te)
$,
respectively, where $(\si,i)=(a,1), (b,2)$,
and the generators $G^{-1}_{\psi }(\te)$, $G^{-1}_S(\te)$, $G^{-1}_{P}(\te)$,
are given in Refs.\cite{Blasone:1995zc}-\cite{Blasone:2001du}, \cite{Capolupo:2004pt,Capolupo:2004}.

In a similar way, the flavor
annihilation operators are written in terms of free fields annihilators, $\al^{r}_{{\bf k},i}$, $b_{{ \bf k},i}$ and $c_{{ \bf k},i}$   as $\al^{r}_{{\bf k},\si}
\equiv G^{-1}_{\psi}(\te)\;\al^{r}_{{\bf k},i} \;G_{\psi}(\te)$,
   $ b_{{ \bf k},\si}\equiv
 G^{-1}_S(\te)\; b_{{ \bf k},i}\;
G_S(\te),$ and $ c_{{ \bf k},\si} \equiv
 G^{-1}_{P}(\te)\; c_{{ \bf k},i}\;
G_{P}(\te).$  They annihilate the flavor vacuum (in the following we suppress the explicit time label)
$|0\ran_{f}\,\equiv\,|0\ran_{f}^{\psi}\, \otimes \,|0\ran_{f}^S\,
\otimes \,|0\ran_{f}^{P}\, $, where,
$
|0\ran_{f}^{\psi}\,\equiv \, G^{-1}_{\psi }(\te) \; |0\ran_{m}^{\psi}\,,
$
$
|0\ran_{f}^S\,\equiv \, G^{-1}_S(\te) \; |0\ran_{m}^S\,,
$ and
$
|0\ran_{f}^{P}\,\equiv \, G^{-1}_{P}(\te) \; |0\ran_{m}^{P}\,,
$
are the flavor vacua of the fields $\psi_{\si}(x)$, $S_{\si}(x)$, $P_{\si}(x)$,
respectively and $|0\ran_{m}^{\psi}$, $|0\ran_{m}^S$, $|0\ran_{m}^{P}$ denote the vacua for free fields.

As before, the vacuum $|0\ran_{f}$ is a  condensate. We have in fact \cite{Capolupo:2007hy}:
$
 {}_{f}\langle 0| \al_{{\bf k},i}^{r \dag} \al^r_{{\bf
k},i} |0\ran_{f}   =   \sin^{2}\te ~ |V^{\psi}_{{\bf
k}}|^{2}\,,
$
and
$
{}_{f}\langle 0| b_{{\bf k},i}^{ \dag} b_{{\bf
k},i} |0\ran_{f}   =     {}_{f}\langle 0| c_{{\bf k},i}^{\dag}
 c_{{\bf k},i} |0\ran_{f} \,=\, \sin^{2}\te ~ |V^B_{{\bf
k}}|^{2}\,,
$
where $i=1,2$ and the reference frame in which ${\bf k}=(0,0,|{\bf k}|)$ has been adopted for convenience.
The  Bogoliubov coefficients  $V^{\psi}_{{\bf k}}$ and  $V^B_{{\bf k}}$ are
\bea\label{Bogoliubov}
|V^{\psi}_{{\bf k}}|  =  \frac{ (\om_{k,1}+m_{1}) - (\om_{k,2}+m_{2})}{2
\sqrt{\om_{k,1}\om_{k,2}(\om_{k,1}+m_{1})(\om_{k,2}+m_{2})}}\, |{\bf k}| \, ,
\qquad
| V^B_{{\bf k}}|  = \frac{1}{2} \lf( \sqrt{\frac{\om_{k,1}}{\om_{k,2}}} -
\sqrt{\frac{\om_{k,2}}{\om_{k,1}}} \ri)\,.
\eea

The expectation value of the fermionic and bosonic parts of $H$ are given by:
$
{}_{f}\lan
0| H_{\psi} | 0 \ran_{f}\, =-\,  \int d^{3}{\bf k} \,
(\omega_{k,1} + \omega_{k,2}) \,(1 - 2\,|V^{\psi}_{\bf k}|^{2}
\sin^{2}\theta)\, ,
$
and
$
 {}_{f}\lan
0| H_B | 0 \ran_{f}\, =\,  \int d^{3}{\bf k} \,
(\omega_{k,1} + \omega_{k,2}) \,(1 + 2\,|V^B_{\bf k}|^{2}
\sin^{2}\theta)\,,
$ respectively.
Then  we have \cite{Capolupo:2010ek}
\bea\label{violation}
\, {}_{f}\lan 0| (H_{\psi}\,+\, H_B) | 0 \ran_{f}\,=\,
2\,\sin^{2}\theta \, \int d^{3}{\bf k} \,
(\omega_{k,1} + \omega_{k,2}) \,(|V^{\psi}_{\bf k}|^{2} + \,|V^B_{\bf k}|^{2})
\,,
\eea
which is different from zero and positive when $\theta\neq0$ and $m_1\neq m_2$. Eq.(\ref{violation}) shows that SUSY breaking is induced by particle mixing phenomenon. This situation should show up in any supersymmetric model in which mixing is turned on before SUSY is spontaneously broken by some other mechanism.

\section{Discussion and Conclusions}

We have shown that, in a supersymmetric field theory, different phenomena may spontaneously break SUSY because of the presence of vacuum condensates. For many of such systems the nontrivial dynamics makes the naive vacuum of the theory unstable in favor of a new vacuum which contains the vacuum condensates \footnote{The reader familiar with superconductivity will undoubtly recognize the basic mechanism of BCS theory in this description.}. The real vacuum of the system is obtained from the original one by applying an appropriate Bogoliubov transformation.
Besides BCS theory, this is expected to happen in the mixing case as well. In fact, while particle mixing is put by hand in the standard model or in neutrino physics, it is expected to result from some dynamical effect which is still unknown, whose result will be a shift of vacuum. In other situations, such as the Unruh or the Schwinger effects, the system is put in an external field. This makes the original vacuum unstable again (Schwinger effect), or even inaccessible to the observer (Unruh effect), in favor of a new vacuum which is again obtained by means of a Bogoliubov transformation controlled by the external field. In the remaining case, that is TFD, the Bogoliubov transformed vacuum is called thermal vacuum, and it is the appropriate one to use at finite temperature in place of the naive one which is valid at zero temperature.
The discussion then boils down to the point that, in any situation in which the physics is described by a vacuum with condensates, SUSY is spontaneously broken.
We have given evidence of SUSY breaking in the case of the free Wess--Zumino model, however our result can be extended to more complex and physically relevant situations. In particular, a well established feature of the Standard Model, namely flavor mixing, emerges as a possible trigger of spontaneous SUSY breaking.

\end{document}